\documentclass{PoS}

\title{Simulations of the Disk-Jet Interaction in GRS~1915+105 and Other Systems}

\ShortTitle{Simulations of the Disk-Jet Interaction}

\author{\speaker{David Rothstein}%
         \thanks{NSF Astronomy and Astrophysics Postdoctoral Fellow.}\\
        Cornell University, Ithaca NY, USA\\
        E-mail: \email{dmr37@cornell.edu}}

\abstract{After an X-ray binary experiences a transient jet ejection, it undergoes a phase in which its X-ray light curve is dominated, for some time, by thermal emission from an accretion disk surrounding the black hole.  The accretion physics in the thermal-dominant state is understood better than in any other, and it is therefore the best state for comparing observations to theoretical models.  Here, I present simulations that study the way a thermally-emitting disk might be expected to behave immediately \emph{after} a large-scale, steady jet has been removed from the system in the form of a sudden ejection.  I simulate the ejection's effect on the disk by allowing the strength of turbulence (modeled by the $\alpha$ parameter of Shakura and Sunyaev) to increase rapidly in time, and I show how this change can lead to an outburst in an otherwise-steady disk.
The motivation for treating the jet removal in this way is the fact that many models for jets involve large-scale magnetic fields that should inhibit the magnetorotational instability believed to drive turbulence; this should naturally lead to a rapid increase in turbulence when the magnetic field is ejected from the system or otherwise destroyed during the ejection event.  I show how the timescale and luminosity of the outburst can be controlled by the manner in which $\alpha$ is allowed to change, and I briefly discuss ways in which these simulations can be compared to observations of X-ray binaries, in particular GRS~1915+105, which shows the most complex and variable behavior of any black hole system in outburst.}

\FullConference{VI Microquasar Workshop: Microquasars and Beyond\\
                 September 18-22, 2006\\
                 Como, Italy}

\begin{document}

\section{Introduction}

Thanks to hard work by many researchers, much is now known observationally about the behavior of X-ray binary systems as they undergo an X-ray outburst coupled with a transient radio jet ejection (e.g., \cite{Corbel2004,FenderUnified2004,Belloni2005,HomanBelloni2005,McClintock2006,RemillardMcClintock2006}).  The evolution during the outburst can be conveniently plotted on a hardness-intensity diagram, showing that the ejections occur as the accretion disk moves from a bright hard to a bright soft X-ray state and as a steady jet structure becomes unobservable at radio wavelengths \cite{FenderUnified2004} (although there is a large amount of variation in the detailed tracks that systems take on this diagram; see, e.g., \cite{RemillardMcClintock2006}).

The soft state, in which thermal emission from the accretion disk is dominant (hence the alternative name ``thermal-dominant state'') and there is no observational evidence for a jet, is the state that is by far the best understood physically.  Going back to the original work of Shakura and Sunyaev \cite{SS73} (who explained the basic structure of a thin disk), Abramowicz and collaborators \cite{Abramowicz1988} (who extended the theory to include moderately thick disks and showed the possibility of limit-cycle oscillations in the inner disk associated with the radiation pressure instability) and many others, the basic physics of the disk during this state is reasonably well-known.  In trying to compare theory to observations, the thermal-dominant state provides the best hope for a stringent test.

Given the observational picture that has been built up, it is worthwhile to ask the following previously-unasked question:  What would we expect to happen to a thermal-dominant accretion disk \emph{immediately after} a large-scale, steady jet has been ejected from its environment?

\begin{figure}
\centering
\includegraphics[width=0.55\textwidth]{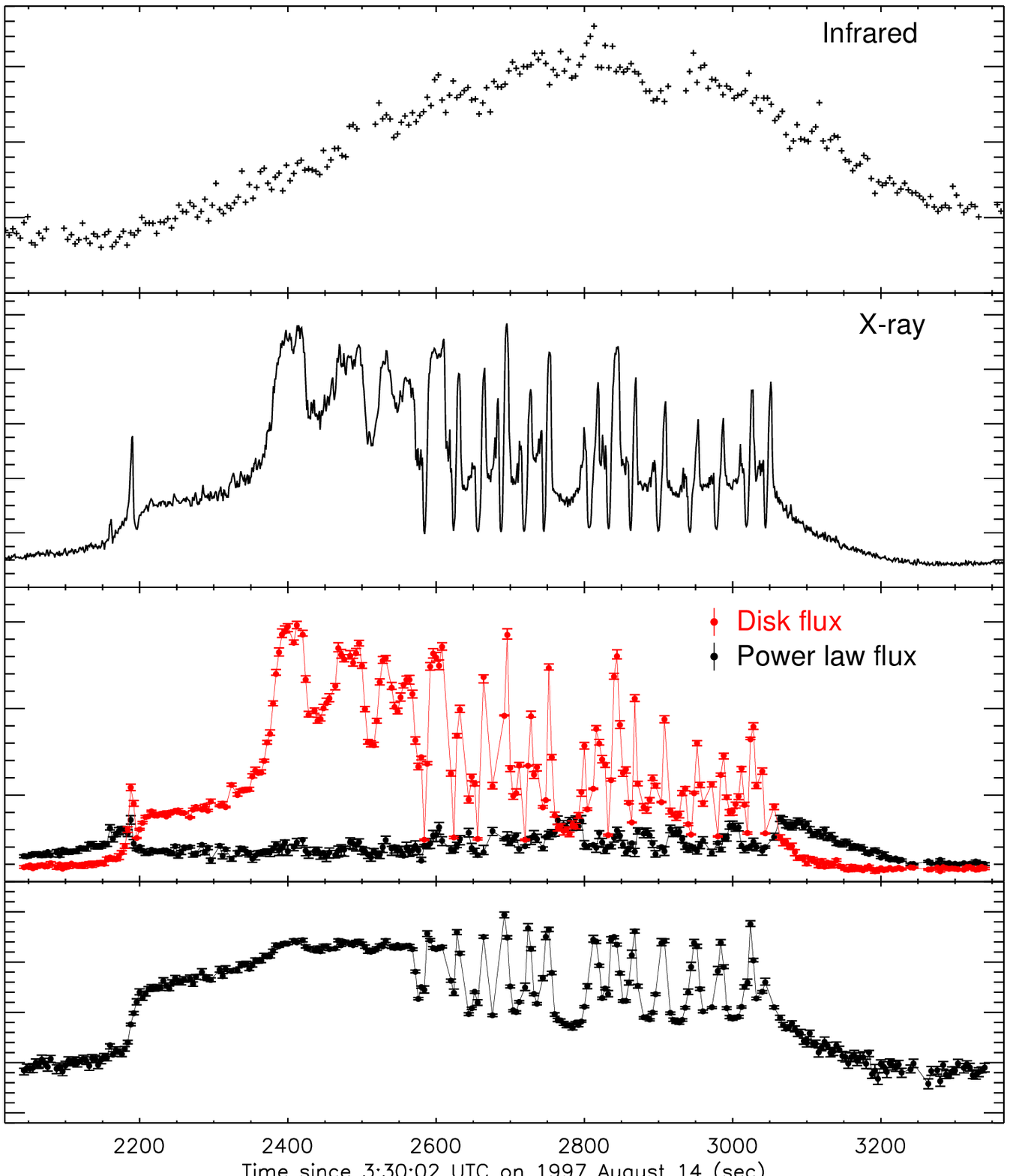}
\caption{An example of a transient jet ejection in GRS~1915+105 on a timescale of $\sim 20$ minutes, based on observations using the {\it Rossi X-ray Timing Explorer} and the Palomar 5 meter telescope \cite{Eikenberry1998}.  The top two panels show the infrared (dereddened by 3.3 magnitudes) and X-ray
light curves, respectively.  The bottom two panels show the results of X-ray spectral fitting at 4-second resolution.  Notice that the infrared flare (signaling a jet ejection) appears to begin around the time of a ``spike'' in the X-ray light curve that corresponds to a state transition, in which thermal emission from the disk begins to dominate over nonthermal power law emission.  Notice also that the behavior of the disk immediately after the spike has a different character than the subsequent pattern of X-ray oscillations, which show rapid changes in the disk temperature.  This figure is modified from one originally presented by Rothstein et al. 2005 \cite{Rothstein2005}.}
\label{fig:grs}
\end{figure}

GRS~1915+105 is the X-ray binary which is best suited to answering this kind of question.  It has essentially spent all of its time since its discovery (nearly
fifteen
years ago) in an ``active'' state, cycling right around the region of the hardness-intensity diagram where transient jets form (e.g., \cite{FenderUnified2004}).  What makes GRS~1915+105 unique is likely its enormous accretion disk, which is much bigger than that of any other similar system and therefore effectively gives the black hole's strong gravity a lot of mass that it can continually ``play with'' \cite{Done2004,FenderBelloni2004,RemillardMcClintock2006}.  Multiwavelength observations of GRS~1915+105 in the X-rays, infrared and radio show that the system sometimes produces repeated ejections on timescales of $\sim 30$ minutes or faster \cite{PooleyFender1997,Eikenberry1998,Mirabel1998,Rothstein2005}.  An example of this behavior is shown in Figure \ref{fig:grs}.  A hard-to-soft state transition occurs around the time of a ``spike'' in the X-ray light curve (at $\sim 2,200$ seconds), after which the nonthermal power law component of the X-ray spectrum becomes weaker and the thermal disk component becomes stronger.  The spike also seems to coincide with the beginning of an infrared flare, signaling that an ejection has taken place.  The timescale over which the state transition happens is extremely rapid, on the order of a couple seconds in this case \cite{Rothstein2005}.

Following the ejection, the disk enters a period of repeating oscillations.  Numerical simulations can generally reproduce the basic temporal \cite{Honma1991} and spectral \cite{Watarai2003} properties of these types of oscillations using a classic limit-cycle instability.  Nonetheless, it is unclear how a limit-cycle instability can be used to explain the \emph{initial} outburst in Figure \ref{fig:grs}, which is clearly of longer duration and a different character than the subsequent oscillations.  Nor is it clear what causes the oscillations to start and the outburst to occur in the first place.  Here, I present a simplified model in which the rapid removal of magnetic field associated with a steady jet can actually \emph{drive} an otherwise-steady disk into an outburst whose properties and timescale can be more easily matched to the initial post-ejection behavior seen in GRS~1915+105 and other objects.

\section{Simulations}

I present results from one-dimensional (i.e., axisymmetric and height-integrated) simulations of a standard accretion disk around a 14 $M_{\odot}$ black hole, appropriate for GRS~1915+105 \cite{GreinerMass2001,Harlaftis2004}.  Radiative and advective cooling are included in the equations, turbulence is treated using the dimensionless $\alpha$ parameter of Shakura and Sunyaev \cite{SS73}, gravity is treated using a pseudo-Newtonian potential \cite{PaczWiita1980}, and large-scale, ordered magnetic fields are ignored (because they are assumed to be removed from the system when the steady jet disappears).  More details of the simulations will be presented in a forthcoming work, but in general, the physics behind them is similar to that included in other one-dimensional disk simulations that have been presented in the literature.

All of the simulations begin with a gas pressure dominated disk with a dimensionless accretion rate $\dot{M} = 0.1$ (measured in units of the Eddington luminosity divided by the speed of light squared).  For these parameters, the entire disk in the region of the simulation (typically $7 r_{g}$ out to a few hundred $r_{g}$, where $r_{g}$ is the gravitational radius) is initially stable.

For the purpose of these simulations, the primary effect of the ``removal of a jet'' is postulated to be a rapid increase in $\alpha$, the parameter that controls the strength of turbulence in the disk.  From an observational standpoint, this idea is attractive because $\alpha$ is essentially the ``clock'' in a standard disk; the thermal and viscous timescales are inversely proportional to it.  Observations such as those in Figure \ref{fig:grs} clearly show that the disk experiences a state transition on a very rapid timescale, and after the transition the typical evolution timescales of the disk are faster.  From a theoretical standpoint, meanwhile, many models suggest, usually indirectly, that the value of $\alpha$ should be different when a jet is and is not present.  These models invoke a picture for a steady jet in which a large-scale, ordered magnetic field is present that can channel matter and energy away from the accretion disk (e.g., \cite{BlandfordPayne1982,vanPutten2003,Livio2003,Tagger2004}).  If these magnetic fields are stronger than equipartition, the magnetorotational instability that is thought to drive accretion is not generally expected to operate \cite{BalbusHawley1991}, and the strength of turbulence (parameterized by $\alpha$) will be weaker.  Tagger and collaborators \cite{Tagger2004} have specifically argued that an outburst such as that shown in Figure \ref{fig:grs} involves the destruction of a large-scale magnetic field via a reconnection event, during which the magnetic field eventually becomes weak enough that the magnetorotational instability can turn on.

The simulations presented here begin with a value of $\alpha=0.01$, which is subsequently increased to $\alpha=0.1$ to mimic the effect of a jet ejection that removes the large-scale magnetic field in a particular region of the disk.  In general, simulations of the magnetorotational instability suggest that $\alpha \sim 0.1$ is an appropriate value when the instability is operating \cite{Hawley2000,Blackman2006}.  The qualitative behavior discussed in the next section is not sensitive to the exact value of $\alpha$ chosen, however.

It is important to point out the difference between this work and the wide variety of other models in the literature where $\alpha$ is allowed to vary in response to other disk parameters (e.g., \cite{MenouNarayan2000,Lasota2001,Truss2004} for X-ray binary outbursts).  The primary difference is that this work explores the effects of a \emph{rapid} change in $\alpha$ due to an external event.  As will be shown, the rapid nature of the change is a key ingredient that can drive an otherwise-stable disk into outburst.  ``Rapid'' in this case means that $\alpha$ must change much faster than the disk is able to respond thermally, or $v_{B} \gg \alpha c_{s}$, where $v_{B}$ is the speed at which the magnetic field is removed vertically from the disk (thereby leading to the change in $\alpha$) and $c_{s}$ is the sound speed.  This should not be difficult to achieve in a magnetically-dominated disk, where the Alfv\'{e}n speed is greater than the sound speed.

\section{Results}

\begin{figure}
\centering
\includegraphics[width=0.6\textwidth]{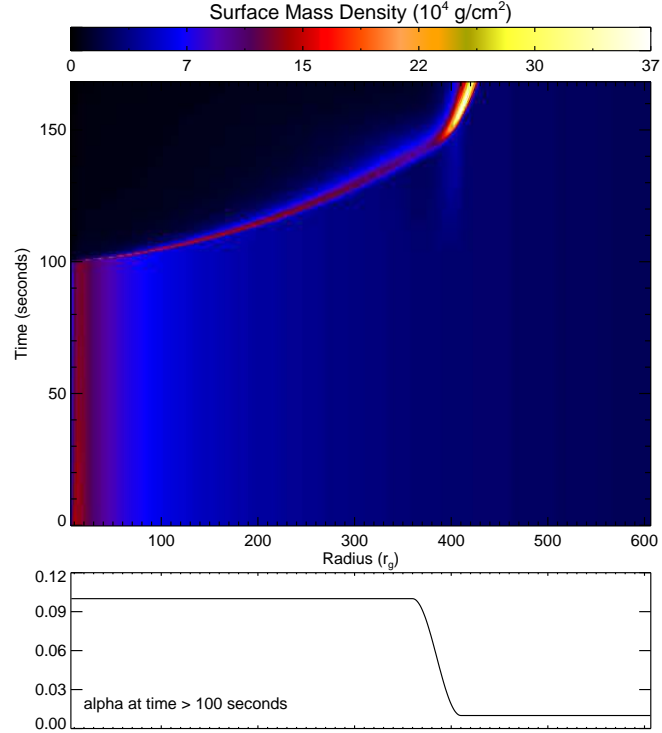}
\caption{The surface mass density evolution for a run in which the value of $\alpha$ is raised from 0.01 to 0.1 in the inner region of the disk ($r \lesssim 400 r_{g}$; bottom panel).  An outburst is initiated in the inner region, largely evacuating the disk and sending a wave of mass outwards, which eventually stalls when it reaches the region where $\alpha$ is still equal to 0.01.}
\label{fig:sigma3d_inner}
\end{figure}

Figure \ref{fig:sigma3d_inner} shows the surface mass density evolution for a run in which the disk was initially held in a steady state with $\alpha=0.01$ for the first 100 seconds, and then $\alpha$ was increased to a value of 0.1 in the inner part of the disk.  The change in $\alpha$ drives the disk into outburst; the inner disk becomes nearly evacuated of matter, and a density wave propagates outward before eventually stalling when it reaches the region where $\alpha$ is still $\sim 0.01$.

\begin{figure}
\centering
\includegraphics[width=0.68\textwidth]{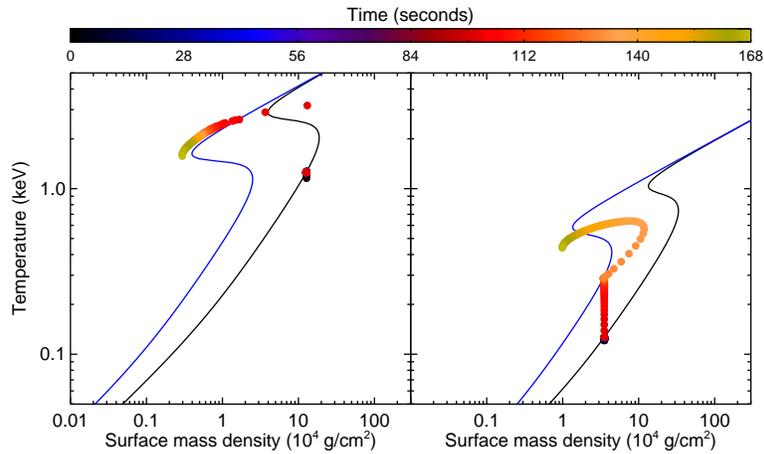}
\caption{The evolution on the local energy balance curve at two radii, when $\alpha$ is raised from 0.01 to 0.1 in the inner region of the disk.  The left panel shows the evolution at $r=22 r_{g}$; the energy balance curve for $\alpha=0.01$ is shown in black, and that for $\alpha=0.1$ is shown in blue.  When $\alpha$ changes, the material at this location in the disk is suddenly in an unstable region of parameter space, and an outburst is triggered.  At a radius of $r=327 r_{g}$, meanwhile (right panel), the disk heats up when $\alpha$ changes, but it still remains on a stable part of the energy balance curve.  Only when the density wave from the inner part of the disk reaches this radius (at $\sim 140$ seconds) is enough mass added to drive it into an outburst.}
\label{fig:scurve_inner}
\end{figure}

Figure \ref{fig:scurve_inner} shows the evolution of the disk on a plot of surface mass density vs. temperature.  The local energy balance curves for $\alpha = 0.01$ and $\alpha = 0.1$ are shown; the classic ``S shape'' of these curves is due to the interplay of radiation pressure and advective cooling \cite{Abramowicz1988}.  We can see from this figure that although the disk is initially stable (and would remain so forever if $\alpha$ did not change), an outburst is triggered in the inner part of the disk simply because the change in $\alpha$ puts the disk in a region of parameter space (with respect to the new energy balance curve) where it can heat up towards the advection-dominated upper branch.  Figure \ref{fig:scurve_inner} also shows that the outburst has a different character further out in the disk; there, the disk heats up when $\alpha$ is increased but does not immediately become unstable.  Instead, the instability is triggered when the density wave arrives and the extra mass pushes this region of the disk ``over the edge.'' 

\begin{figure}
\centering
\includegraphics[width=0.65\textwidth]{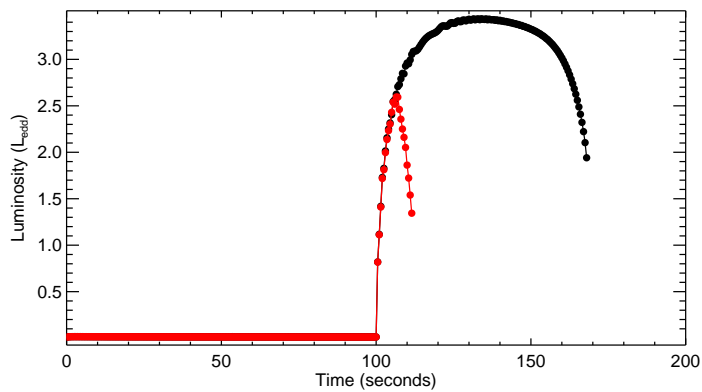}
\caption{The outburst light curve for the run shown in Figures 2 and 3 (black), where the value of $\alpha$ is raised for $r \lesssim 400 r_{g}$, compared to the outburst light curve for a run in which the value of $\alpha$ is only raised for $r \lesssim 100 r_{g}$ (red).  The luminosity is plotted in units of the Eddington luminosity.}
\label{fig:light_curve_inner}
\end{figure}

The outburst light curve is shown in Figure \ref{fig:light_curve_inner} (black points). We see that the disk remains bright as long as the density wave is propagating through the disk and supplying mass to trigger new outbursts at large radii and simultaneously sustain the original outburst at the inner radii (cf. Figure \ref{fig:sigma3d_inner}).  When the density wave stalls, however, the inner radii stop receiving mass, and eventually the outburst dies down as the disk is able to cool.

The density wave feature seen in these simulations is not new; it was observed in some of the earliest one-dimensional simulations of accretion disk outbursts (e.g., \cite{Honma1991}) and in many others works since then.  In those simulations, outbursts in an $\alpha=0.1$ disk were driven not by a change in $\alpha$, but rather by an external accretion rate $\dot{M}$ that was high enough so that the equilibrium state of the disk would put it in the unstable portion of the local energy balance curve.  In these simulations, the density wave stalls at a radius smaller than $\sim 150-200 r_{g}$, leading to an outburst of duration $\sim 20-30$ seconds \cite{Honma1991,SzuszMiller1998,SzuszMiller2001}.

In the simulations I present here, there are several important differences.  The timescale and radial extent of the outburst are not fixed, but rather depend on the size of the region in which $\alpha$ is changed.  It is possible to obtain an outburst that is much longer than normal, as in the case discussed above, or much shorter, which will occur if we only allow $\alpha$ to change over a small region of the disk (as shown by the red points in Figure \ref{fig:light_curve_inner}).  Within a large range of radii over which $\alpha$ might be allowed to change, the density wave will always stall near the outer boundary of the region where $\alpha$ is altered (i.e., the outer boundary of the region where the ``jet ejection'' takes place).  The reason that changing $\alpha$ can give longer outbursts than the original accretion disk simulations can be appreciated by looking at the right panel of Figure \ref{fig:scurve_inner}; by increasing $\alpha$ rapidly, we effectively obtain an $\alpha=0.1$ disk that has the surface mass density of an $\alpha=0.01$ disk.
At any given radius, the disk is very close to being near the critical bend in the energy balance curve at which it becomes unstable, and the density wave only needs to add a small amount of mass to trigger a new outburst at that radius.

These results have important observational implications.  The original simulations of \cite{Honma1991} and others predicted continuously repeating outburst cycles, because the disks they studied were inherently unstable.  GRS~1915+105, however (as well as other black hole candidates), frequently undergoes an outburst after a period of quiescence.  The simulations presented here can reproduce this behavior easily, because it is the change in $\alpha$ that drives an {\it otherwise stable disk} into an outburst.  Furthermore, if the value of $\dot{M}$ is such that the disk is stable for both the initial and final values of $\alpha$, then we would expect to see a single outburst, with the disk returning to a stable state on the ``new'' energy balance curve after the outburst is complete.  For a certain range of $\dot{M}$, however, it is possible to have the disk be stable for the low value of $\alpha$ but unstable for the high value.  In this case, the disk would be expected to undergo repeated oscillations following the initial outburst.  Furthermore, these subsequent oscillations might be of the more ``normal'' variety found by \cite{Honma1991} and others, because the disk at this point will be in a normal $\alpha=0.1$ state.  Thus, one could imagine light curves with some resemblance to those shown in Figure \ref{fig:grs}, where the initial outbursts are of a different character and longer duration than the subsequent oscillations.  Unfortunately, numerical instabilities associated with the inner disk boundary currently prevent investigation of the outbursts discussed here over more than one cycle, but I plan to pursue this in a future work.

\begin{figure}
\centering
\includegraphics[width=0.6\textwidth]{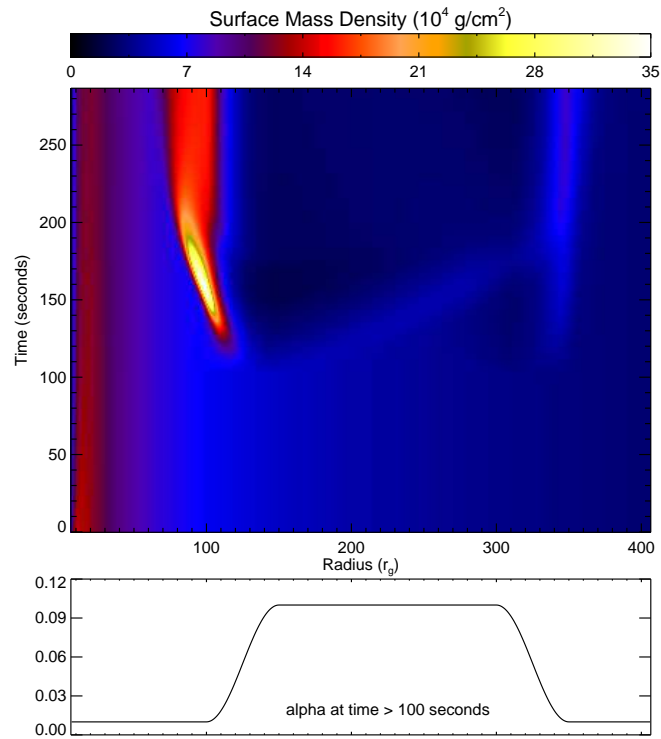}
\caption{The surface mass density evolution for a run in which the value of $\alpha$ is raised from 0.01 to 0.1 in the middle part of the disk (cf. Figure 2).  In this case, mass builds up around the boundary of the region where $\alpha$ is changed and penetrates it slightly, but never reaches the inner part of the disk.  Also, unlike Figure 2, there is no prominent density wave initiated when $\alpha$ changes.}
\label{fig:sigma3d_middle}
\end{figure}

Figures \ref{fig:sigma3d_middle} shows a run in which the value of $\alpha$ is only changed in the middle part of the disk, not the inner part.  In this case, the change in $\alpha$ causes mass to build up at $r \sim 100 r_{g}$ without penetrating to the extreme inner region.  Examination of the local energy balance curves in this case shows that enough mass eventually builds up to trigger an outburst, but unlike in the case discussed above, this outburst is triggered by the existence of a gradient in $\alpha$ within the disk, not the speed with which $\alpha$ changes.

\begin{figure}
\centering
\includegraphics[width=0.65\textwidth]{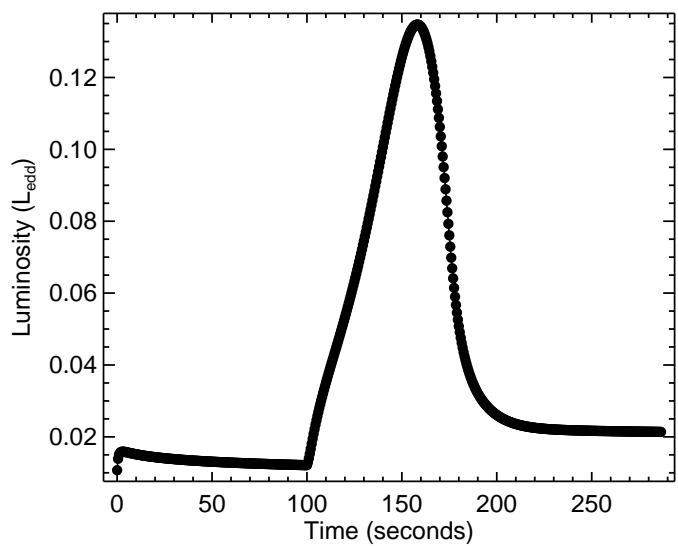}
\caption{The outburst light curve for the run shown in Figure 5.  The outburst peak is significantly delayed from the time ($t = 100$ seconds) at which $\alpha$ was changed.}
\label{fig:light_curve_middle}
\end{figure}

The outburst light curve from this run is shown in Figure \ref{fig:light_curve_middle}.  The peak of the disk outburst is significantly delayed from the ``jet ejection'' (which occurred at $\sim 100$ seconds).  This could be relevant for observations of GRS~1915+105, where infrared activity in the jet sometimes appears to precede significant X-ray activity from the disk (see Figure \ref{fig:grs}, or observations by Eikenberry and collaborators \cite{Eikenberry2000} for a more extreme example).  These results can also be compared to the scenario investigated by \cite{Lovelace1994}, where a disturbance in the magnetic field at large radii led to an ejection as well as significant accretion into the inner disk---Figure \ref{fig:light_curve_middle} shows that when $\alpha$ is allowed to vary, the outburst does not necessarily have to reach the extreme inner disk in order to be observable.

\section{Conclusions}

A rapid increase in the strength of turbulence in an accretion disk (as might be associated with the removal of a large-scale magnetic field during a transient jet ejection) appears to be a promising way to explain how an otherwise-steady disk can be driven into outburst, as well as to explain some of the details of the outburst light curves seen in GRS~1915+105 and other X-ray binaries.  Future work should involve a more rigorous comparison of the model to observations, as well as comparison to X-ray spectra.  Overall, the period of activity in an accretion disk immediately \emph{after} a jet has been ejected may be a promising place to test theoretical models, because the physics of the disk in this phase is well-understood but the behavior of the disk may still be influenced by fingerprints of the jet.

\acknowledgments

I would like to thank Richard Lovelace for his valuable advice and assistance during this work, as well as Steve Eikenberry for many discussions and suggestions.  David Rothstein is supported by an NSF Astronomy and Astrophysics Postdoctoral Fellowship under award AST-0602259.

\bibliography{droth}

\begin{thebibliography}{10}

\bibitem{Abramowicz1988}
M.~A. {Abramowicz}, B.~{Czerny}, J.~P. {Lasota}, and E.~{Szuszkiewicz},
  \emph{{Slim accretion disks}}, \apj\, \textbf{332}, 646--658 (1988).

\bibitem{BalbusHawley1991}
S.~A. {Balbus} and J.~F. {Hawley}, \emph{{A powerful local shear instability in
  weakly magnetized disks. I. Linear analysis}}, \apj\, \textbf{376}, 214--233
  (1991).

\bibitem{Belloni2005}
T.~{Belloni}, J.~{Homan}, P.~{Casella}, M.~{van der Klis}, E.~{Nespoli},
  W.~H.~G. {Lewin}, J.~M. {Miller}, and M.~{M{\'e}ndez}, \emph{{The evolution
  of the timing properties of the black-hole transient GX 339-4 during its
  2002/2003 outburst}}, \aap\, \textbf{440}, 207--222 (2005) [{\tt
  astro-ph/0504577}].

\bibitem{Blackman2006}
E.~G. {Blackman}, R.~F. {Penna}, and P.~{Varniere}, preprint
  (astro-ph/0607119), July 2006.

\bibitem{BlandfordPayne1982}
R.~D. {Blandford} and D.~G. {Payne}, \emph{{Hydromagnetic flows from accretion
  discs and the production of radio jets}}, \mnras\, \textbf{199}, 883--903
  (1982).

\bibitem{Corbel2004}
S.~{Corbel}, R.~P. {Fender}, J.~A. {Tomsick}, A.~K. {Tzioumis}, and
  S.~{Tingay}, \emph{{On the Origin of Radio Emission in the X-Ray States of
  XTE J1650-500 during the 2001-2002 Outburst}}, \apj\, \textbf{617},
  1272--1283 (2004) [{\tt astro-ph/0409154}].

\bibitem{Done2004}
C.~{Done}, G.~{Wardzi{\'n}ski}, and M.~{Gierli{\'n}ski}, \emph{{GRS 1915+105:
  the brightest Galactic black hole}}, \mnras\, \textbf{349}, 393--403 (2004)
  [{\tt astro-ph/0308536}].

\bibitem{Eikenberry1998}
S.~S. {Eikenberry}, K.~{Matthews}, E.~H. {Morgan}, R.~A. {Remillard}, and R.~W.
  {Nelson}, \emph{{Evidence for a Disk-Jet Interaction in the Microquasar GRS
  1915+105}}, \apjl\, \textbf{494}, L61--L64 (1998) [{\tt astro-ph/9710374}].

\bibitem{Eikenberry2000}
S.~S. {Eikenberry}, K.~{Matthews}, M.~{Muno}, P.~R. {Blanco}, E.~H. {Morgan},
  and R.~A. {Remillard}, \emph{{Faint Infrared Flares from the Microquasar GRS
  1915+105}}, \apjl\, \textbf{532}, L33--L36 (2000) [{\tt astro-ph/0001472}].

\bibitem{vanPutten2003}
S.~S. {Eikenberry} and M.~H.~H.~M. {van Putten}, preprint (astro-ph/0304386),
  April 2003.

\bibitem{FenderBelloni2004}
R.~P. {Fender} and T.~{Belloni}, \emph{{GRS 1915+105 and the Disc-Jet Coupling
  in Accreting Black Hole Systems}}, \araa\, \textbf{42}, 317--364 (2004) [{\tt
  astro-ph/0406483}].

\bibitem{FenderUnified2004}
R.~P. {Fender}, T.~M. {Belloni}, and E.~{Gallo}, \emph{{Towards a unified model
  for black hole X-ray binary jets}}, \mnras\, \textbf{355}, 1105--1118 (2004)
  [{\tt astro-ph/0409360}].

\bibitem{GreinerMass2001}
J.~{Greiner}, J.~G. {Cuby}, and M.~J. {McCaughrean}, \emph{{An unusually
  massive stellar black hole in the Galaxy}}, \nat\, \textbf{414}, 522--525
  (2001) [{\tt astro-ph/0111538}].

\bibitem{Harlaftis2004}
E.~T. {Harlaftis} and J.~{Greiner}, \emph{{The rotational broadening and the
  mass of the donor star of GRS 1915+105}}, \aap\, \textbf{414}, L13--L16
  (2004) [{\tt astro-ph/0312373}].

\bibitem{Hawley2000}
J.~F. {Hawley}, \emph{{Global Magnetohydrodynamical Simulations of Accretion
  Tori}}, \apj\, \textbf{528}, 462--479 (2000) [{\tt astro-ph/9907385}].

\bibitem{HomanBelloni2005}
J.~{Homan} and T.~{Belloni}, \emph{{The Evolution of Black Hole States}},
  \apss\, \textbf{300}, 107--117 (2005) [{\tt astro-ph/0412597}].

\bibitem{Honma1991}
F.~{Honma}, S.~{Kato}, and R.~{Matsumoto}, \emph{{Nonlinear oscillations of
  thermally unstable slim accretion disks around a neutron star or a black
  hole}}, \pasj\, \textbf{43}, 147--168 (1991).

\bibitem{Lasota2001}
J.-P. {Lasota}, \emph{{The disc instability model of dwarf novae and low-mass
  X-ray binary transients}}, New Astronomy Review\, \textbf{45}, 449--508
  (2001) [{\tt astro-ph/0102072}].

\bibitem{Livio2003}
M.~{Livio}, J.~E. {Pringle}, and A.~R. {King}, \emph{{The Disk-Jet Connection
  in Microquasars and Active Galactic Nuclei}}, \apj\, \textbf{593}, 184--188
  (2003) [{\tt astro-ph/0304367}].

\bibitem{Lovelace1994}
R.~V.~E. {Lovelace}, M.~M. {Romanova}, and W.~I. {Newman}, \emph{{Implosive
  accretion and outbursts of active galactic nuclei}}, \apj\, \textbf{437},
  136--143 (1994).

\bibitem{McClintock2006}
J.~E. {McClintock} and R.~A. {Remillard}, \emph{{Black hole binaries}}, Compact
  Stellar X-ray Sources (Walter {Lewin} and Michiel {van der Klis}, eds.),
  Cambridge University Press, Cambridge, 2006, p.~157.

\bibitem{MenouNarayan2000}
K.~{Menou}, J.-M. {Hameury}, J.-P. {Lasota}, and R.~{Narayan}, \emph{{Disc
  instability models for X-ray transients: evidence for evaporation and low
  {$\alpha$}-viscosity?}}, \mnras\, \textbf{314}, 498--510 (2000) [{\tt
  astro-ph/0001203}].

\bibitem{Mirabel1998}
I.~F. {Mirabel}, V.~{Dhawan}, S.~{Chaty}, L.~F. {Rodr{\'{\i}}guez},
  J.~{Mart{\'{\i}}}, C.~R. {Robinson}, J.~{Swank}, and T.~{Geballe},
  \emph{{Accretion instabilities and jet formation in GRS 1915+105}}, \aap\,
  \textbf{330}, L9--L12 (1998) [{\tt astro-ph/9711097}].

\bibitem{PaczWiita1980}
B.~{Paczy{\'n}ski} and P.~J. {Wiita}, \emph{{Thick accretion disks and
  supercritical luminosities}}, \aap\, \textbf{88}, 23--31 (1980).

\bibitem{PooleyFender1997}
G.~G. {Pooley} and R.~P. {Fender}, \emph{{The variable radio emission from GRS
  1915+105}}, \mnras\, \textbf{292}, 925--933 (1997) [{\tt astro-ph/9708171}].

\bibitem{RemillardMcClintock2006}
R.~A. {Remillard} and J.~E. {McClintock}, \emph{{X-Ray Properties of Black-Hole
  Binaries}}, \araa\, \textbf{44}, 49--92 (2006) [{\tt astro-ph/0606352}].

\bibitem{Rothstein2005}
D.~M. {Rothstein}, S.~S. {Eikenberry}, and K.~{Matthews}, \emph{{Observations
  of Rapid Disk-Jet Interaction in the Microquasar GRS 1915+105}}, \apj\,
  \textbf{626}, 991--1005 (2005) [{\tt astro-ph/0501624}].

\bibitem{SS73}
N.~I. {Shakura} and R.~A. {Sunyaev}, \emph{{Black holes in binary systems.
  Observational appearance.}}, \aap\, \textbf{24}, 337--355 (1973).

\bibitem{SzuszMiller1998}
E.~{Szuszkiewicz} and J.~C. {Miller}, \emph{{Limit-Cycle Behaviour of Thermally
  Unstable Accretion Flows on to Black Holes}}, \mnras\, \textbf{298}, 888--896
  (1998) [{\tt astro-ph/9804233}].

\bibitem{SzuszMiller2001}
E.~{Szuszkiewicz} and J.~C. {Miller}, \emph{{Non-linear evolution of thermally
  unstable slim accretion discs with a diffusive form of viscosity}}, \mnras\,
  \textbf{328}, 36--44 (2001) [{\tt astro-ph/0107257}].

\bibitem{Tagger2004}
M.~{Tagger}, P.~{Varni{\`e}re}, J.~{Rodriguez}, and R.~{Pellat},
  \emph{{Magnetic Floods: A Scenario for the Variability of the Microquasar GRS
  1915+105}}, \apj\, \textbf{607}, 410--419 (2004) [{\tt astro-ph/0401539}].

\bibitem{Truss2004}
M.~R. {Truss} and G.~A. {Wynn}, \emph{{Long time-scale variability in GRS
  1915+105}}, \mnras\, \textbf{353}, 1048--1054 (2004) [{\tt
  astro-ph/0406499}].

\bibitem{Watarai2003}
K.~{Watarai} and S.~{Mineshige}, \emph{{Model for Relaxation Oscillations of a
  Luminous Accretion Disk in GRS 1915+105: Variable Inner Edge}}, \apj\,
  \textbf{596}, 421--428 (2003) [{\tt astro-ph/0306548}].

\end{thebibliography}
\bibliographystyle{PoS}

\end{document}